**Tuning valleys and wave functions of van der Waals heterostructures by varying the number of layers: A first-principles study**

*Muhammad S. Ramzan, Jens Kunstmann and Agnieszka B. Kuc\**

Muhammad S. Ramzan, Dr. Agnieszka B. Kuc
Department of Physics and Earth Sciences, Jacobs University Bremen, Campus Ring 1, 28759 Bremen, Germany, Helmholtz-Zentrum Dresden-Rossendorf, Abteilung Ressourcenökologie, Forschungsstelle Leipzig, Permoserstr. 15, 04318 Leipzig, Germany
E-mail: a.kuc@hzdr.de

Dr. Jens Kunstmann
Theoretical Chemistry, TU Dresden, 01062 Dresden, Germany

In van der Waals heterostructures of two-dimensional transition-metal dichalcogenides (2D TMDCs) electron and hole states are spatially localized in different layers forming long-lived interlayer excitons. Here, we have investigated, from first principles, the influence of additional electron or hole layers on the electronic properties of a $MoS_2/WSe_2$ heterobilayer (HBL), which is a direct band gap material. Additional layers modify the interlayer hybridization, mostly affecting the quasiparticle energy and real-space extend of hole states at the $\Gamma$ and electron states at the $Q$ valleys. For a sufficient number of additional layers, the band edges move from $K$ to $Q$ or $\Gamma$, respectively. Adding electron layers to the HBL leads to more delocalized $Q$ states, while $\Gamma$ states do not extend much beyond the HBL, even when more hole layers are added. These results suggest a simple and yet powerful way to tune band edges and the real-space extend of the electron and hole wave function in TMDC heterostructures, strongly affecting the lifetime and dynamics of interlayer excitons.

# 1. Introduction

Two-dimensional (2D) transition-metal dichalcogenides (TMDCs) have been intensively studied in the past decade, not only as single layers,[1,2] but, more recently, also as bilayer (2L) van der Waals heterostructures (vdW HS),[3–11] where two different monolayers (1L) are vertically stacked, forming ubiquitous moiré patterns, due to the lattice mismatch and twist angle between individual layers.[12–14] Such vdW HS offer band gaps smaller than these of parental single layers and interlayer excitons, which are longer-lived, because of the type-II band alignment, in which electrons and holes reside in different layers.[15–19] These interlayer excitons have been observed in different TMDC vdW HS with either crystallographic twist angles (0° or 60°)[20,21] or with arbitrary, random stackings.[15,18,22,23] Moreover, 1L TMDCs and vdW HS obey different optical selection rules.[24,25] Thus, one can speak of new physical phenomena in these materials and they are being considered for a wide range of applications, ranging from ultrafast charge separation, (opto)electronics through quantum information processing or valleytronics.[17,19,26–31]

A large variety of 2L vdW HS have recently been investigated on the basis of experiments and theory,[18,22,31–33] including these constituting of layers from Group 6 TMDCs.[5–7,11,16,34–36] For instance, Kim et al.,[19] reported ultra-long valley lifetime in $MoS_2/WSe_2$ heterostructures, which was explained by the spatial confinement of electron and hole in different layers. Park et al.[33] investigated the gate response of such vdW HS and reported ambipolar or unipolar transport behavior depending on the extent of charge depletion in the heterojunction. The key parameter, the interlayer coupling, which predominantly dictates the properties of vdW HS, can be tuned, e.g., via twist angle between the layers or by careful selection of individual layers. For instance, Tongay et al.[22] demonstrated that the interlayer coupling of TMDC vdW HS can be tuned by vacuum thermal annealing, where light emission spectrum gradually evolves from single layers, contributing separately to the novel coupled spectrum as function of the interlayer distance.

Up to date, the majority of research focuses on vdW HS made of just two different TMDC layers. However, band structure engineering by systematically extending the number of layers of the heterostructure has been much less explored.[37,38] Prominent questions to be addressed are: (i) Can we exploit interlayer coupling by mixing different number of layers to tune the multi-valley, electronic structure near the band edges? (ii) Will additional electron or hole layers delocalize or localize the interlayer excitons in the vdW HS? (iii) For how many layers of the multi-stack, would the band gap be direct or indirect?

Here, in order to address these questions, we investigate the influence of additional electron or hole layers (nL, with n = 1–4) on the electronic properties of a $MoS_2/WSe_2$ heterobilayer (HBL). Our results show that adding additional $WSe_2$ layers to the HBL (forming 1L-$MoS_2$/(n+1)L-$WSe_2$ HS) will mostly affect the valence (hole) states, while adding $MoS_2$ layers to the HBL (forming (n+1)L-$MoS_2$/1L-$WSe_2$ HS) will affect the conduction (electron) states. We find that the heterostructures of up to 1L-$MoS_2$/4L-$WSe_2$ or 3L-$MoS_2$/1L-$WSe_2$ stay direct gap semiconductors (at the $K$ point), while more layers in both HS would open new transition channels at the $Q$ and/or $\Gamma$ points. We, therefore, provide the energy differences between $k$ valleys in the conduction ($Q$-$K$) and the valence ($K$-$\Gamma$) bands, values which can be obtained experimentally, e.g., from carrier dynamics measurements.[21] The calculated partial charge densities (PCD) for the electron and hole states, analyzed for each relevant $k$ point of the band edges, are either delocalized throughout all the layers or localized in the heterolayer, depending on the stoichiometry of vdW HS. Here, we apply a simple two-step approach of analyzing electron and hole states in large incommensurate HS using single unit cells, where we minimize the effect of the tensile strain or compression, resulting from the differences in the lattice vectors of constituent layers.

## 2. Results and Discussion

First, electronic of bulk and 1L $MoS_2$ and $WSe_2$ were obtained (see **Figure S1** in Supporting Information (SI)). Good agreement with previous works supports the choice of our simulation methods.[39,40] While 1L-MoS$_2$ is a direct band gap semiconductor (at K), 1L WSe$_2$ is an indirect gap material,[41] with the corresponding direct band gap at K being only 50 meV higher in energy. Both 1L systems exhibit spin splitting in the valence band at the K point of 197 meV and 606 meV for MoS$_2$ and WSe$_2$, respectively, due to strong SOC. It is well known that HSE06 overestimates the spin split for TMDCs.[42]

Next, we have calculated band structures of vdW HS, HBL/nL-MoS$_2$ and HBL/nL-WSe$_2$ (with n = 1–4; see **Figure 1**), using the fully relaxed structures (method I) and our proposed 2-step approach (method II) for comparison (see Computational Details in Section 4 for detailed explanations on employed methods). In method II, we perform two different calculations to analyze the electron and hole states separately. By using different lattice constants for the electron and hole calculations, our approach removes the problematic tensile strain or compression, which arises from the incommensurability of the layers. This approach is valid for our systems, because they are type-II HS. Because with method I both $H_h^h$ and $R_h^M$ stackings result in very similar electronic structures (see **Figure S2** to **Figure S5** in SI) and energy differences, we have only used the $H_h^h$ systems for method II. Here, we first shortly discuss the results from approach I and then more detailed results are shown for the approach II.

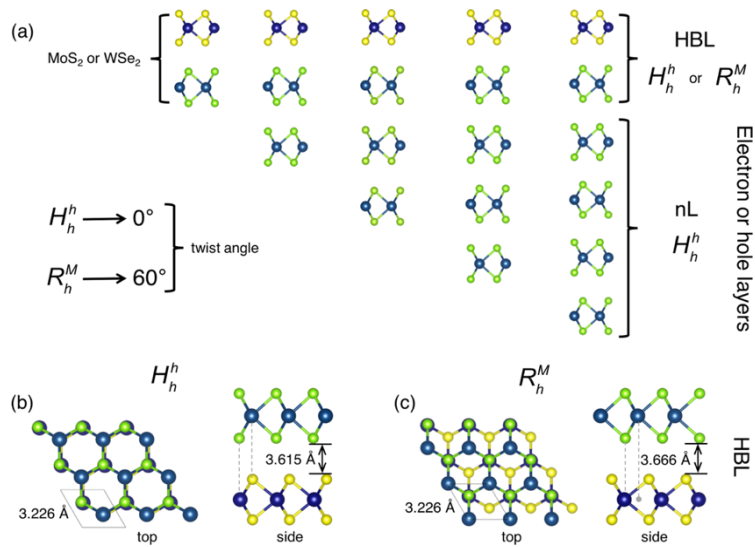

**Figure 1.** The van der Waals heterostructures studied in this work: (a) A MoS$_2$/WSe$_2$ heterobilayer (HBL) with additional electron or hole layers (nL = 1–4) of either MoS$_2$ or WSe$_2$, respectively. The stacking polytype of the HBL is $H_h^h$ or $R_h^M$ and the one of the nL and its interface with HBL is $H_h^h$. (b and c) Top and side views of the considered stacking polytypes ($H_h^h$, also known as 2H, and $R_h^M$, also known as 3R) in HBLs with 0° and 60° twist angles.

While stacking results in very similar band structures, more significant differences are observed for the multi-stacks: 1) the band gap increases (decreases) for HBL/nL-MoS$_2$ (HBL/nL-WSe$_2$); 2) when n is varied for HBL/nL-MoS$_2$ (HBL/nL-WSe$_2$), the conduction (valence) bands are mostly affected; 3) for HBL/nL-MoS$_2$ (HBL/nL-WSe$_2$), the valley at the $Q$ point in the CB (at the $\Gamma$ point in the VB) decreases (increases) in energy with number of layers nL. These trends are valid for both methods I and II. The variations are noticeable when discussing the energy differences and the direct vs. indirect band gap characters. To understand these differences, we used method I and calculated atom-resolved band structures (so-called fatbands), which are shown in **Figure 2**. Both systems are type-II HS, meaning that the valence band maximum (VBM) is composed mostly of the WSe$_2$ states and conduction band minimum (CBM) is dominated by the MoS$_2$ states. Both HS have direct band gaps at the $K$ point.

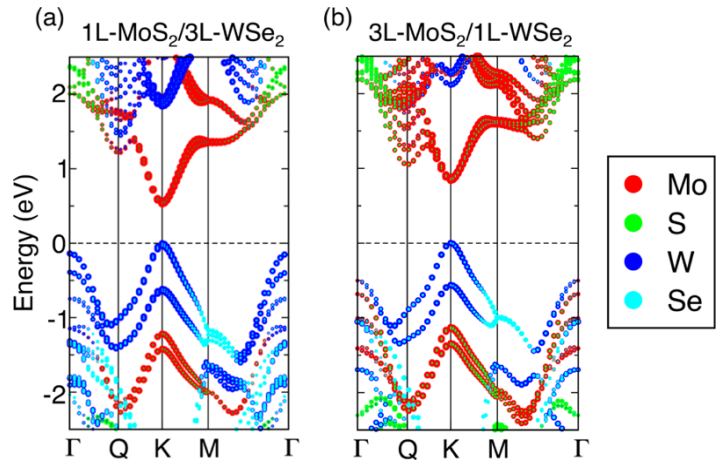

**Figure 2.** The type-II band alignment of the considered van der Waals heterostructures, as shown by atom-resolved, qualitative band structures; the color-code corresponds to the atom's character. The conduction band edge is dominated by MoS$_2$ states and the valence band edge by WSe$_2$ states. The two band structures are only of qualitative nature, because the lattice mismatch between MoS$_2$ and WSe$_2$ is ignored in this plot. The Fermi energy is indicated by dashed horizontal lines and was shifted to the valence band maximum. The corresponding in-plane lattice parameters (a = b) of relaxed vdW HS are 3.257 Å and 3.196 Å for (a) and (b), respectively.

The exemplary band structures from method II are shown in **Figures 3 and 4** for 3L-MoS$_2$/1L-WSe$_2$ and 1L-MoS$_2$/3L-WSe$_2$, respectively. Depending on which part of the heterostructure is fully relaxed, 1L or nL, the respective VB or CB are taken into consideration (see Section 4 for details). We have again analyzed the orbital resolved band structure, which consistently show that all HS are of type-II.

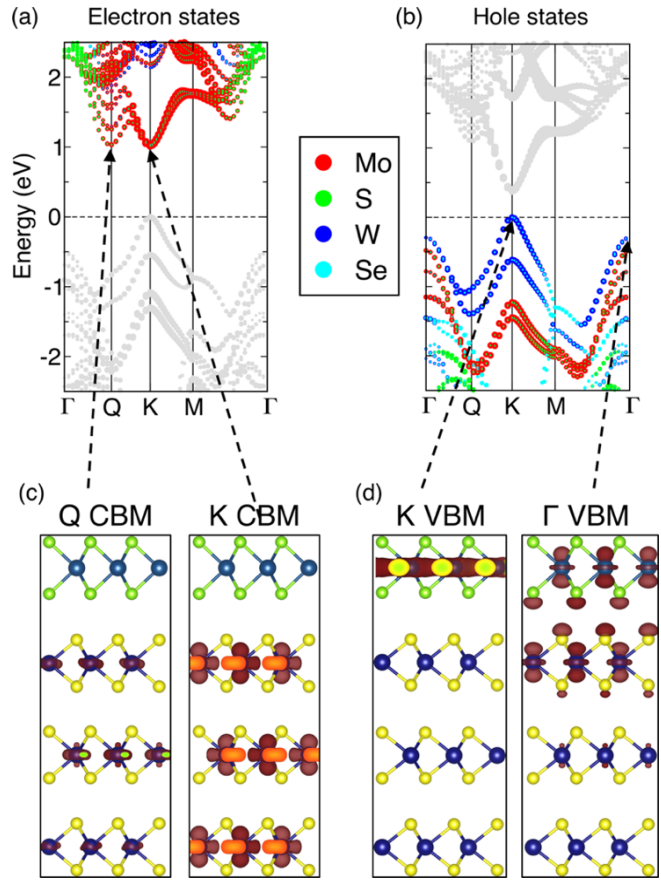

**Figure 3.** Electron and hole states of 3L-$MoS_2$/1L-$WSe_2$ at different valleys and their wave function extend in real space. (a, b) Atom-resolved band structures, where the color-code corresponds to the atom character. (c, d) Isosurfaces of the partial charge density (iso value = 0.1) of indicated states, illustrating very different "wave function extends". While the electron states at K and Q are delocalized over the whole $MoS_2$ stack, the hole state at K is localized only in the $WSe_2$ layer and the hybridized Γ point hole state is essentially localized in the $MoS_2$/$WSe_2$ heterobilayer. Note, in c) at the K point in CBM, the energy difference between the three states is only 10 meV, thus, they are plotted together and regarded as degenerate. Different lattice constants (a = b), 3.167 Å and 3.294 Å, are used for the calculations in (a, c) and (b, d), respectively, to properly represent the electron or the hole states of the lattice-mismatched heterostructure.

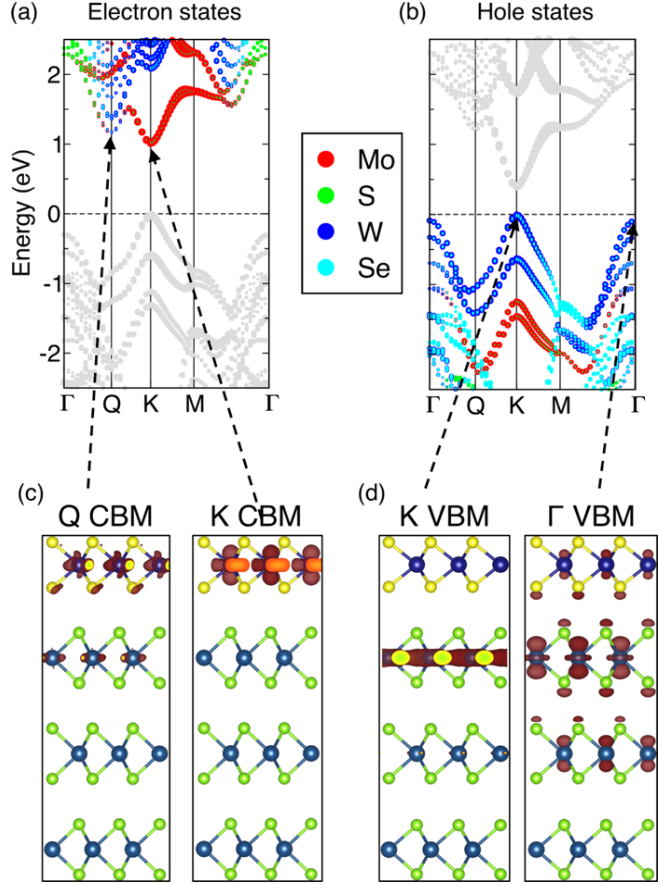

**Figure 4.** Electron and hole states of 1L-MoS$_2$/3L-WSe$_2$ at different valleys and their wave function extend in real space. (a, b) Atom-resolved band structures, where the color-code corresponds to the atom character. (c, d) Isosurfaces of the partial charge density (iso value = 0.1) of indicated states, illustrating very different "wave function extends". The electron states at Q and the hole state at Γ are partially hybridized between MoS$_2$ and WSe$_2$. Interestingly, the K point hole state almost does not extend beyond the first WSe$_2$ layer. This is, because the energy differences between VBM and VBM-1 or VBM-2 are 19 meV and 28 meV, respectively, thus, it is regarded as non-degenerate. Different lattice constants (a = b), 3.166 Å and 3.287 Å, are used for the calculations in (a, c) and (b, d), respectively, to properly represent the electron or the hole states of the lattice-mismatched heterostructure.

The most significant difference between approach I and II is how fast the $Q$ valley in the CB ($\Gamma$ valley in the VB) decreases (increases) in energy. In other words, the energy differences between the $Q$ and $K$ valleys in the CB and the $\Gamma$ and $K$ valleys in the VB, indicating direct or indirect band gaps, differ for both approaches (see **Figure 5**). Approach I would indicate that both types of HS with at least 6 layers in the multi-stack are still direct-gap materials. However, approach II shows that these changes happen much faster and already 4L-MoS$_2$/1L-WSe$_2$ and 1L-MoS$_2$/5L-WSe$_2$ transform to indirect-gap materials.

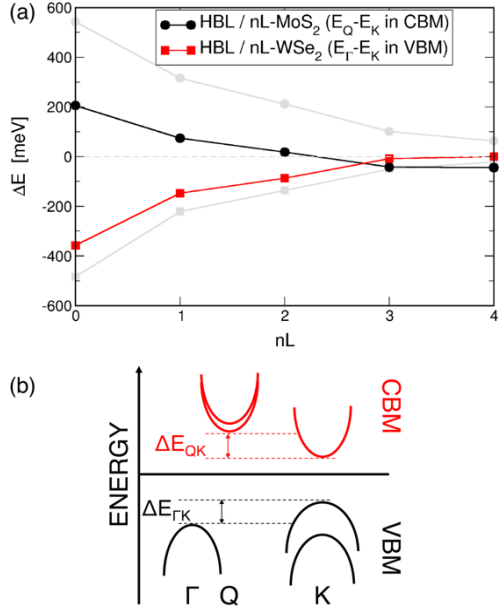

**Figure 5.** (a) Direct to indirect band gap transition when adding electron or hole layers to the heterobilayer (HBL). The plot shows the energy differences between the K and the Q valley $\Delta E_{QK}$ for electrons and between the K and the Γ valley ($\Delta E_{\Gamma K}$) for holes (see legend). Gray lines are results for method I, with strained unit cells. Points above (black) or below (red) zero indicate a direct-gap, e.g., HBL/3L-MoS$_2$ has an indirect band gap. Note that n = 0 corresponds to HBL. (b) A generic band structure focused on selected k-points showing $\Delta E_{QK}$ and $\Delta E_{\Gamma K}$.

Additionally, we have calculated the PCD of electron and hole states (cf. **Figure 3** and **4**). We selected the band edge states at four relevant *k*-points, i.e., the *K*- and *Q*-valley for the electron states and the *K*- and *Γ*-valley for the hole states. The type-II character is also shown by the PCD plots at the *K* point, where the states are localized in the respective building blocks of the HS, meaning in the 1HL and the (n+1)L stack. Similar results were reported by Kunstmann et al.,[52] for MoS$_2$/WSe$_2$ HBL. Both the *Γ* and the *Q* valley states are mixtures of *d*-orbitals of metal and *p*-orbitals of the chalcogen atoms, while states at *K* are dominated by *d*-orbitals of the metal atoms. There is a vanishing overlap of the states in the *K* valley, which are separated by at least 6.5 Å (in the bilayer case) in a purely interlayer exciton. Additional HL has a different influence on the localization of electron or hole states in the (n+1)L: while 1L-WSe$_2$ results in a nearly equal distribution of electron states (at *Q* and *K*) in all three MoS$_2$ layers (in 3L-MoS$_2$/1L-WSe$_2$), see **Figure 3c**, 1L-MoS$_2$ does not, and the largest contribution of hole states in case of 1L-MoS$_2$/3L-WSe$_2$ is in the WSe$_2$ layer closest to the 1L-MoS$_2$, and it reduces when going further down in layers (see **Figure 4d**).

In other words, the states at the $\Gamma$ point are essentially localized in the HBL, when more electron layers are added, and in the HBL and the closest to it WSe$_2$ layer, when hole layers are added. This also suggests that adding more layers to the nL does not significantly affect the overlap at $\Gamma$. On the other hand, adding more layers to the nL-MoS$_2$ results in stronger delocalization of the electron states at $Q$ point, while it does not affect these states when more layers in nL-WSe$_2$ are added. The level of (de)localization of states will affect the lifetime and intensity of the resulting interlayer excitons.

Furthermore, we can also conclude, from **Figure 5**, that changing the number of layers in nL results mostly in the changes in positions of the $Q$ and the $\Gamma$ points: the former is reducing and the latter increasing in energy when increasing nL. This suggests that, even though the materials are direct-gap semiconductors with transitions at the $K$ point from these DFT simulations, a prominent indirect-transitions between $\Gamma-K/Q$ and $K-Q$ points should also be present, meaning the materials would have layer-indirect and momentum-indirect excitons, what would result in longer lifetimes.[21]

In present work, we avoid discussing the absolute band gap values. Our 2-step approach can also be used to analyze band gap values when the band edges are calculated with respect to the vacuum energy. From approach I, we have noticed that in HS with nL-MoS$_2$, the band gaps increase, while in HS based on nL-WSe$_2$, the band gaps decreased with n (cf. **Figure S2** to **S6**). This subject, however, requires more detailed investigations and is beyond the scope of this work.

## 3. Conclusion

We have investigated the influence of additional electron or hole layers (nL, with n = 1–4) on the electronic properties of a MoS$_2$/WSe$_2$ heterobilayer (HBL). We showed that this approach can tune the valleys and the wave functions of the multi-stack in different ways, depending on the stoichiometry of the layers. We have studied two types of systems, namely (n+1)L-$MoS_2$/1L-$WSe_2$ and 1L-$MoS_2$/(n+1)L-$WSe_2$. Furthermore, we compared a very common approach, where the heterostructure is modelled with small, but strongly strained, unit cells with our simple 2-step approach, where two calculations are performed to analyze the electron and hole states separately. By using different lattice constants for the electron and hole calculations, our approach removes the problematic tensile strain or compression, which arises from the incommensurability of the layers. It can be used, because the HS are type-II materials, where the electrons reside predominantly in the $MoS_2$ layers and holes predominantly in $WSe_2$ layers.

We have analyzed the energy differences between the band edges at the *K* valley and other band edge valleys, i.e., the *Q* valley in the conduction and the *Γ* valley in the valence band. These energy differences can be obtained experimentally, e.g., from carrier dynamics measurements. We show that increasing the number of additional layers, the *Q-K* and *Γ-K* energy differences decrease and increase in energy, respectively. Our calculations suggest that the materials stay direct-gap semiconductors for up to 3L-$MoS_2$/1L-$WSe_2$ and 1L-$MoS_2$/4L-$WSe_2$. From partial-charge density calculations of the electron and hole states, we could show that both types of HS differ significantly in the state localization: when increasing the number of $MoS_2$ layers, the CB states at *Q* and *K* are delocalized more or less equally throughout all $MoS_2$ layers. The VBM states at *K* are completely localized in the 1L-$WSe_2$, while at *Γ*, it is practically localized in the HBL. This means that *Q* point offers stronger delocalization when increasing the number of layers, while no significant change is observed for the *Γ* states. In case of changing the nL of $WSe_2$, the *Q* and *K* states in CB (localized to 1L-$MoS_2$) and *K* in VB (localized to the $WSe_2$ closest to the 1L-$MoS_2$) are almost unaffected by increasing nL, and *Γ* is more or less localized to three layers (1L-$MoS_2$ and two closest $WSe_2$ layers). All of these states are almost unaffected by the number of layers in the multi-stack of $WSe_2$. This results in HS with different localization of states, thus, resulting in interlayer excitons of shorter or longer lifetimes. These differences may also affect the exciton dynamics.

Our study offers a simple experimental approach to understand and control excitons in TMDC heterostructures.

## 4. Experimental Section

*Computational Details*: All simulations were performed using density functional theory (DFT) methods as implemented in the Vienna *ab-initio* simulation package (VASP).[39] The projector-augmented wave (PAW) method was used to describe the interactions between electrons and nuclei. The exchange and correlation potentials were treated using the generalized-gradient approximation (GGA)[40] functional proposed by Perdew, Burke, and Ernzerhof (PBE) for full geometry optimization (lattice vectors and atomic positions). Plane-wave cutoff of 500 eV was used in all simulations. The D3 dispersion correction was included, as proposed by Grimme.[41] A vacuum layer of at least 15 Å was added along the out-of-plane direction to avoid spurious interactions with the next periodic images. All structures were relaxed until all the forces acting on each atom were less than $2\times10^{-2}$ eV Å$^{-1}$ and the total energy change between two self-consistent steps was less than $1\times10^{-4}$ eV. A $\Gamma$-centered $12\times12\times1$ *k*-point mesh was used to sample the Brillouin zone during structural optimizations. It is well known that standard GGA DFT functionals underestimated the electronic band gaps, thus, electronic properties were calculated using the Heyd-Scuseria-Ernzerhof hybrid (HSE06)[42,43] functional, considering relativistic effects, such as spin-orbit coupling (SOC). Even though, we do not discuss the band gap sizes in the present work, we have also noticed strong deviations in the shape of bands in band structures between PBE and HSE06, thus, we used the latter for the electronic structure analysis. The PCD was, however, calculated using PBE with SOC and analyzed for exemplary structures, i.e., (HBL + 2L) 4L vdW HS. Structural illustrations and PCD plots were created using the 3D visualization package VESTA.[44]

The systems studied in this work are shown in **Figure 1**. We consider a MoSe$_2$/WSe$_2$ HBL and systematically add one to four additional layers (nL) of either WSe$_2$ or MoS$_2$, creating either a 1L-MoS$_2$/(n+1)L-WSe$_2$ or a (n+1)L-MoS$_2$/1L-WSe$_2$ vdW HS. The (n+1)L multilayers were kept in their most stable stacking arrangement, $H_h^h$ (also known as 2H). The interface between MoS$_2$ and WSe$_2$ is realized either in the $H_h^h$ or $R_h^M$ (also known as 3R) stacking fashion, which are among the most stable high-symmetry stacking arrangement observed in 0° and 60° twist angle moiré structures.[25] Other stacking polytypes, e.g., $H_h^M$ or $R_h^h$, are, of course, also possible,[21,25] but beyond the scope of the present work, because the electronic properties are affected much stronger by the number of layers than the stacking itself.

A lattice mismatch between the building blocks of HS results in an incommensurate moiré superstructure that, in the best case, would need to be approximated by supercells containing hundreds or thousands of atoms (see **Figure S6**). Such large systems are, however, too demanding for DFT calculations, especially when including SOC with hybrid functionals. The most commonly used approach is to relax such a HS in a small unit cell (referred to as method I in the text below), however, this results in strain or compression of the constituent layers, which is an unrealistic representation of the HS. As previously shown in different theoretical and experimental works, tensile strain and compression strongly affect electronic structures of TMDC layers.[45,46] Thus here, we propose a simple 2-step approach (referred to as method II) to separately analyze the electron and hole states of an incommensurate HS by performing two calculations within primitive unit cells. We suggest that, in order to analyze the electron states near the Fermi level, we use a lattice constant (*a*) of the relaxed (n+1)L-MoS$_2$, while for the hole states, we use *a* of the relaxed (n+1)L-WSe$_2$. This is possible, because of the type-II band alignment, which is discussed further below. We used the same approach in our recent work on the W-based HBLs.[47] In Table S1 in SI, we summarized the lattice parameters for (n+1)L-MoS$_2$ and (n+1)L-WSe$_2$, and we compare them to the lattice parameters of fully relaxed HS.

**Supporting Information**

Supporting Information is available below.

**Acknowledgements**

This research was supported by the Deutsche Forschungsgemeinschaft (project GRK 2247/1 (QM3)). The authors acknowledge the high-performance computing center of ZIH Dresden for computational resources.

# Supporting Information

**Tuning valleys and wave functions of van der Waals heterostruc-tures by varying the number of layers: A first-principles study**

*Muhammad S. Ramzan, Jens Kunstmann and Agnieszka B. Kuc\**

**Table S1.** Calculated lattice parameters for monolayers, bulks, and multi-layer stacks of $MoS_2$, $WSe_2$, and heterostructures of either HBL/nL-$MoS_2$ or HBL/nL-$WSe_2$ (with n = 1 – L) from two different approaches (see Computational Details in main text for details). The results are in good agreement with the available experimental reports (3.160 Å and 3.280 Å for $MoS_2$ and $WSe_2$ bulk, respectively).[1]

| System | | $a$ [Å] | System | | $a$ [Å] |
|---|---|---|---|---|---|
| *Approach I (fully relaxed HS)* | | | | | |
| 1L-$MoS_2$/1L-$WSe_2$ | 2H | 3.226 | 1L-$MoS_2$/1L-$WSe_2$ | 2H | 3.226 |
| | 3R | 3.226 | | 3R | 3.226 |
| 1L-$MoS_2$/2L-$WSe_2$ | 2H | 3.248 | 2L-$MoS_2$/1L-$WSe_2$ | 2H | 3.205 |
| | 3R | 3.247 | | 3R | 3.205 |
| 1L-$MoS_2$/3L-$WSe_2$ | 2H | 3.257 | 3L-$MoS_2$/1L-$WSe_2$ | 2H | 3.196 |
| | 3R | 3.257 | | 3R | 3.197 |
| | com | 3.166 | | com | 3.167 |
| | str | 3.287 | | str | 3.294 |
| 1L-$MoS_2$/4L-$WSe_2$ | 2H | 3.26 | 4L-$MoS_2$/1L-$WSe_2$ | 2H | 3.185 |
| | 3R | 3.26 | | 3R | 3.186 |
| 1L-$MoS_2$/5L-$WSe_2$ | 2H | 3.261 | 5L-$MoS_2$/1L-$WSe_2$ | 2H | 3.182 |
| | 3R | 3.264 | | 3R | 3.182 |
| *Approach II (relaxation of HS to corresponding lattice vectors of layers)* | | | | | |
| Hole States | | | Electron States | | |
| 1L-$MoS_2$/1L-$WSe_2$ | 2H | 3.294 | 1L-$MoS_2$/1L-$WSe_2$ | 2H | 3.166 |
| 1L-$MoS_2$/2L-$WSe_2$ | 2H | 3.288 | 2L-$MoS_2$/1L-$WSe_2$ | 2H | 3.164 |
| 1L-$MoS_2$/3L-$WSe_2$ | 2H | 3.287 | 3L-$MoS_2$/1L-$WSe_2$ | 2H | 3.167 |
| 1L-$MoS_2$/4L-$WSe_2$ | 2H | 3.286 | 4L-$MoS_2$/1L-$WSe_2$ | 2H | 3.163 |
| 1L-$MoS_2$/5L-$WSe_2$ | 2H | 3.287 | 5L-$MoS_2$/1L-$WSe_2$ | 2H | 3.161 |
| Bulk $WSe_2$ | 2H | 3.291 | Bulk $MoS_2$ | 2H | 3.163 |

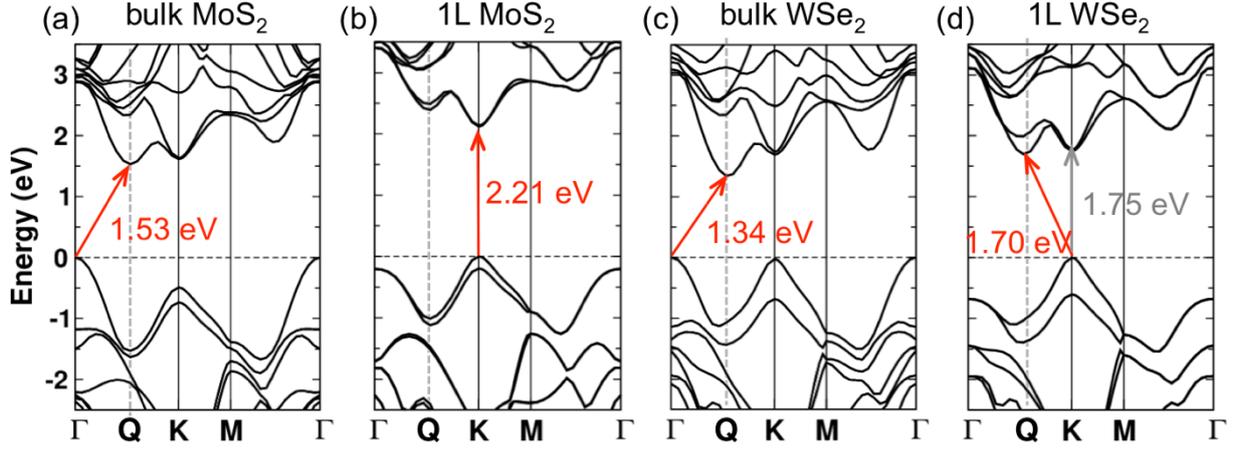

**Figure S1**. Band structure of MoS$_2$ (a, b) and WSe$_2$ (c, d) in bulk (a, c) and monolayer (b, d) forms calculated at the level of HSE06. Both bulk structures are indirect band gap materials with transitions between the $\Gamma$ and $Q$ points. 1L MoS$_2$ is a direct band gap (at the $K$ point) material, while 1L WSe$_2$ is indirect (between $K$ and $Q$ points) band gap system,[2] with the direct gap at the $K$ point being by 50 meV higher in energy.

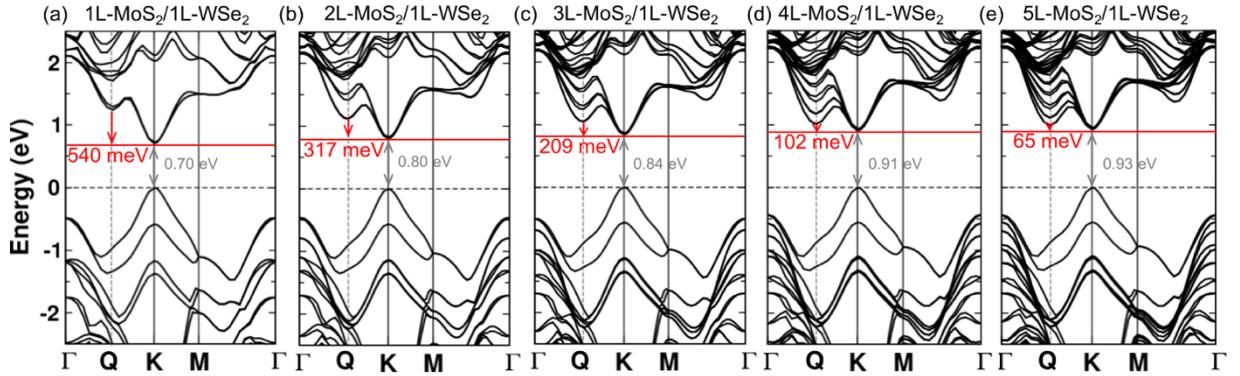

**Figure S2**. Band Structures of (n+1)L-MoS$_2$/1L-WSe$_2$ heterostructures with the $H_h^h$ stacking at the interface. Grey arrows indicate the fundamental band gaps, which are in all cases direct at the $K$ point. Increasing number of layers in MoS$_2$ stack affects mostly the conduction band (CB), with $Q$ point lowering in energy, when increasing number of layers. The valence band (VB) is mostly affected at the $\Gamma$ point, which is, however, much lower in energy than the $K$ point. The difference in energy between the $K$ and $Q$ points in CB is given in red for each system. For convenience, the Fermi level was shifted to VBM at zero and indicated by horizontal dashed lines.

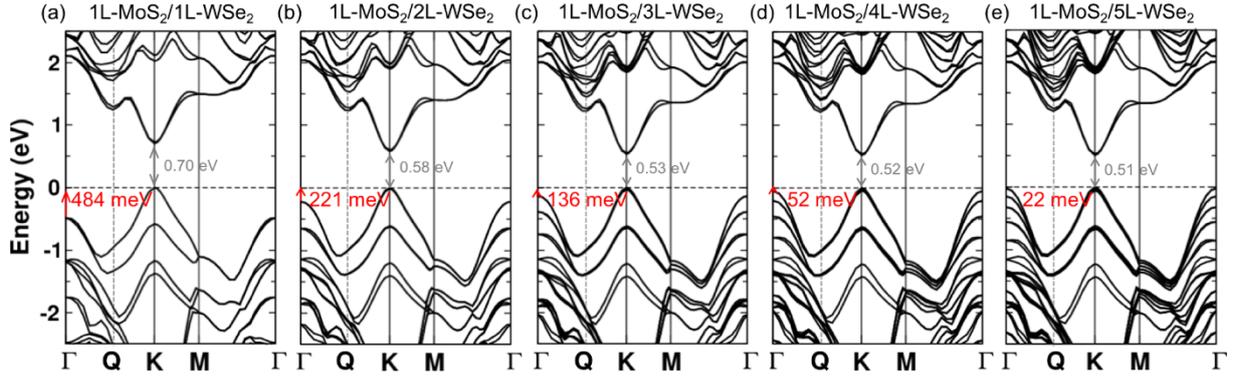

**Figure S3.** Band Structures of 1L-MoS$_2$/(n+1)L-WSe$_2$ heterostructures with the $H_h^h$ stacking at the interface. Grey arrows indicate the fundamental band gaps, which are in all cases direct at the $K$ point. Increasing number of layers in WSe$_2$ stack affects mostly the valence band (VB), with the $\Gamma$ point increasing in energy, when increasing number of layers. The difference in energy between the $K$ and $\Gamma$ points in VB is given in red for each system. For convenience, the Fermi level was shifted to VBM at zero and indicated by horizontal dashed lines.

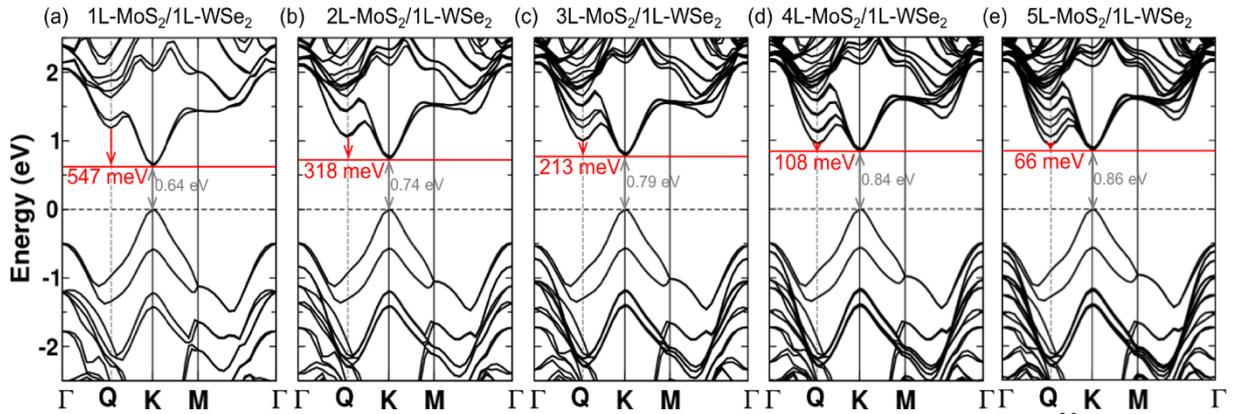

**Figure S4.** Band Structures of (n+1)L-MoS$_2$/1L-WSe$_2$ heterostructures with the $R_h^M$ stacking at the interface. Grey arrows indicate the fundamental band gaps, which are in all cases direct at the $K$ point. Increasing number of layers in MoS$_2$ stack affects mostly the conduction band, with $Q$ point lowering in energy with increasing number of layers. The valence band is mostly affected at the $\Gamma$ point, which is, however, much lower in energy than the $K$ point. The difference in energy between the $K$ and $Q$ points in the conduction band is given in red for each system. For convenience, the Fermi level was shifted to VBM at zero and indicated by horizontal dashed lines

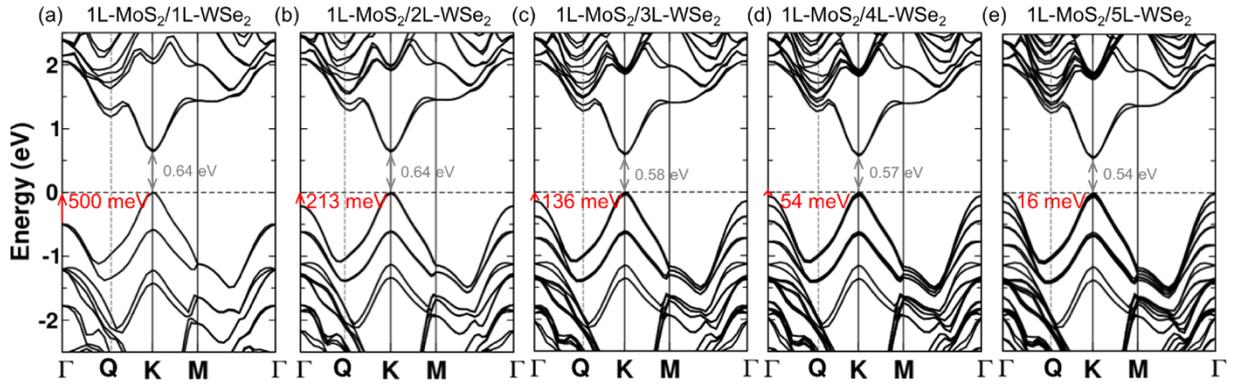

**Figure S5**. Band Structures of 1L-MoS$_2$/(n+1)L-WSe$_2$ heterostructures with the $R_h^M$ stacking at the interface. Grey arrows indicate the fundamental band gaps, which are in all cases direct at the *K* point. Increasing number of layers in WSe$_2$ stack affects mostly the valence band, with *Γ* point increasing in energy with increasing number of layers. The difference in energy between the *K* and *Γ* points in the valence band is given in red for each system. For convenience, the Fermi level was shifted to VBM at zero and indicated by horizontal dashed lines.

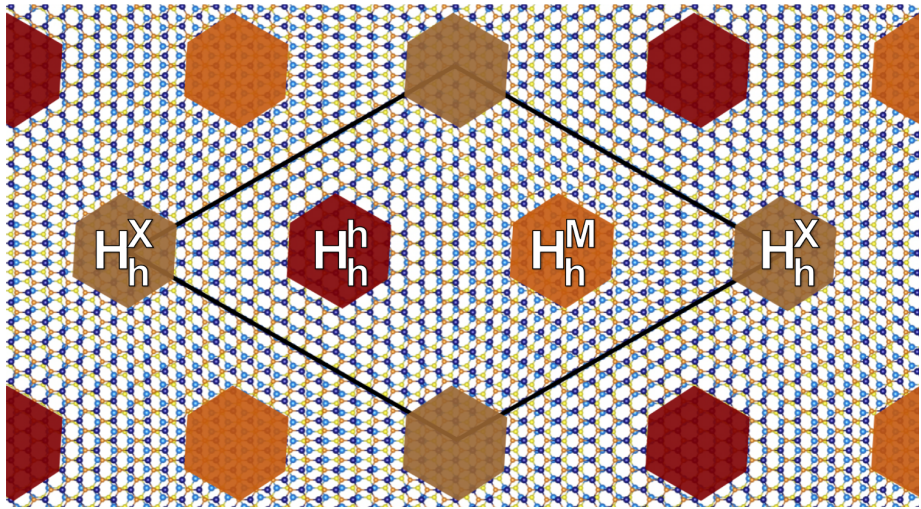

**Figure S6**. Schematic representation of the moiré pattern of the MoS$_2$/WSe$_2$ heterobilayer with the 60° twist angle. Red, orange, and brown hexagons mark regions with high-symmetry stacking configurations. The 0° twist angle results in a similar representation with different high-symmetry stackings. Both require a few thousand of atoms in the supercell to minimize the strain, which is a result of the lattice mismatch between the individual layers.